\documentclass[epj]{svjour}
\usepackage{graphics}
\begin{document}
\title{Topological phase in $1D$ topological Kondo insulator: $Z_{2}$ topological insulator, Haldane-like phase and Kondo breakdown}
\author{Yin Zhong\inst{1}\thanks{\emph{Present address:} zhongy05@hotmail.com} \and Yu Liu\inst{2},\inst{3} \and Hong-Gang Luo \inst{1},\inst{4}
}                     
\institute{Center for Interdisciplinary Studies $\&$ Key Laboratory for
Magnetism and Magnetic Materials of the MoE, Lanzhou University, Lanzhou 730000, China \and LCP, Institute of Applied Physics and Computational Mathematics, Beijing 100088, China \and Software Center for High Performance Numerical Simulation,China Academy of Engineering Physics, Beijing 100088, China \and Beijing Computational Science Research Center, Beijing 100084, China}
\date{Received: date / Revised version: date}
%
\abstract{
We have simulated a half-filled $1D$ $p$-wave periodic Anderson model with numerically exact projector quantum Monte Carlo technique, and the system is indeed located in the Haldane-like state as detected in previous works on the $p$-wave Kondo lattice model, though the soluble non-interacting limit corresponds to the conventional $Z_{2}$ topological insulator.
The site-resolved magnetization in an open boundary system and strange correlator for the periodic boundary have been used to identify the mentioned topological states. Interestingly,
the edge magnetization in the Haldane-like state is not saturated to unit magnetic moment due to the intrinsic charge fluctuation in our periodic Anderson-like model, which is beyond the description of the Kondo lattice-like model in existing literature. The finding here underlies the correlation driven topological state in this prototypical interacting topological state of matter and naive use of non-interacting picture should be taken care.
Moreover, no trace of the surface Kondo breakdown at zero temperature is observed and it is suspected that frustration-like interaction may be crucial in inducing such radical destruction of Kondo screening.
The findings here may be relevant to our understanding of interacting topological materials like topological Kondo insulator candidate SmB$_{6}$.
} 
\maketitle
\section{Introduction}\label{intr}
Recently, the interest in the topological state of matter has been reignited since theoretical prediction and experimental realization
of quantum spin Hall effect and $3D$ topological insulator.\cite{Hasan2010,Qi2011,Hohenadler2013,Bansil2016,Witten2016,Chiu2016,Senthil2015,Liu2016,Dzero2016}
For most of real-life topological materials, for example
time-reversal invariant $Z_{2}$ topological insulator in HgTe/CdTe quantum wells, Bi and Sb-series compounds,\cite{Hasan2010,Qi2011}
their basic properties can be readily understood in the framework of single-particle picture, e.g. topological band theory.\cite{Bansil2016}

However, electron correlation effect in strongly interacting topological materials is still poorly understood, particularly for the well-known topological Kondo
insulator candidate$-$Samarium Hexaboride (SmB$_{6}$).\cite{Dzero2016} Although topological point of view provides appealing explanation to its mysterious low-temperature surface state,\cite{Dzero2016,Dzero2010}
the light electron detected in such surface state hinders a satisfactory solution.\cite{Jiang2013,Neupane2013,Xu2013,Frantzeskakis2013,Xu2014,Li2014} Furthermore, recent high-field quantum oscillation measurement even gives rise to a possibility of a hidden three-dimensional Fermi surface in such bulk insulator.\cite{Tan2015}

In order to answer these serious problems, insightful ideas like surface Kondo breakdown, fractionalized Fermi liquid and Majorana Fermi sea are proposed
in light of the wisdom that strong electron correlation may lead to radical reconstruction of low-energy physics and fractionalization
of electrons is a fascinating option.\cite{Alexandrov2015,Thomson2016,Erten2016,Baskaran2015}

Unfortunately, in dimension larger than one, it is hard to verify these interesting ideas unbiasedly in terms of current analytical and numerical calculation tools, thus
present studies mainly focus on a simplified one-dimensional topological Kondo insulator model, i.e. $1D$ $p$-wave Kondo lattice model invented by
Alexandrov and Coleman, \cite{Alexandrov2014} and further inspected by abelian bosonization and density matrix renormalization group (DMRG) techniques.\cite{Lobos2015,Mezio2015,Hagymasi2016}
The core finding in these works is that if quantum fluctuation effect is included, the paramagnetic end modes (usual fermion zero-energy mode) predicted in large-N mean-field theory are unstable toward a magnetic end state, which is a novel realization of the famous Haldane phase for spin-one antiferromagnetic chain.\cite{Lobos2015,Mezio2015,Hagymasi2016}

Nevertheless, we recall that both the concept of topological Kondo insulator itself and its realistic material modeling start from the periodic Anderson-like model, which includes essential
charge/valence fluctuation effect of $f$-electrons beyond the spin-only Kondo lattice-like model.\cite{Dzero2016,Dzero2010,Alexandrov2013}
For example, x-ray core-level spectroscopy measurement suggests the $f$ level occupation of SmB$_{6}$ is approximated as $0.7$,\cite{Chazalviel1976} much smaller than unit as modeled in $p$-wave Kondo lattice model, thus it is not known whether the results found in such Kondo lattice-like model are applicable to the realistic mixed-valence compound like SmB$_{6}$.
\begin{figure}
\resizebox{0.5\textwidth}{!}{%
  \includegraphics{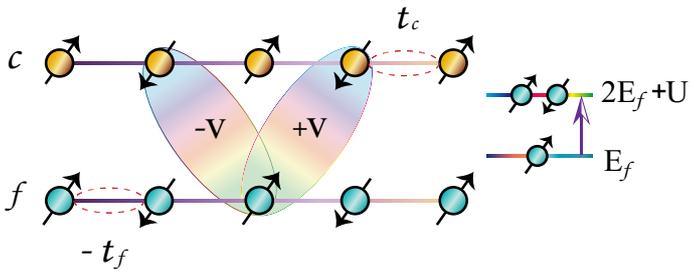}%
}
\caption{$1D$ $p$-wave periodic Anderson model describes a $p$-wave-like hybridization $\pm V$ between local electron orbital (blue) and its neighboring conducting charge carrier (yellow).
The singly occupied local electron has energy $E_{f}$ while double occupation has extra Coloumb energy $U$. The conduction (local) electron hops $t_{c}$ ($-t_{f}$) between nearest-neighbor sites.}
\label{fig:PAM}
\end{figure}

Here, we directly study the $1D$ $p$-wave periodic Anderson model,\cite{Lobos2015,Coleman2015} (See Fig.~\ref{fig:PAM}) which is a natural extension of previous $p$-wave Kondo lattice model.
In this model, the charge degree of freedom of $f$-electron is preserved and not only Kondo limit but also mixed-valence regime can be reliably explored by existing theoretical
approaches. As a first step toward this interesting issue, we consider a half-filled system with particle-hole symmetry ($E_{f}=-U/2$), thus the state of the art projector quantum Monte Carlo (PQMC) can be readily utilized
without the notorious fermion minus-sign problem.\cite{Assaad2008}

\begin{figure}
\resizebox{0.4\textwidth}{!}{%
  \includegraphics{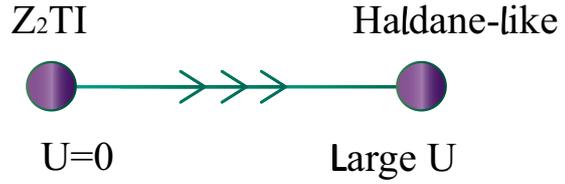}%
}
\caption{The free fixed point of non-interacting $Z_{2}$ topological insulting state ($Z_{2}$ TI) is unstable and flows into strong coupling Haldane-like phase
for $1D$ $p$-wave periodic Anderson model studied in the main text.}
\label{fig:phase}
\end{figure}

From our numerical simulation of PQMC,
it is found that the system is indeed located in the Haldane-like state as detected in previous works on $p$-wave Kondo lattice model, though the soluble non-interacting limit
corresponds to the conventional $Z_{2}$ topological insulating state. (See also Fig.~\ref{fig:phase})
Therefore, our calculation indicates that the finding in Kondo lattice-like models may still be relevant to realistic materials with sensible charge fluctuation at least for nearly half-filled systems.
More interestingly, the edge magnetization in the Haldane-like phase generally deviates from unit magnetic moment as in a pure (spin-one) Haldane phase.
This is due to the intrinsic charge fluctuation introduced in our periodic Anderson-like model and is beyond the description of Kondo lattice-like model in previous works.
Additionally, we have not observed any noticeable signature of surface Kondo breakdown at least for the zero temperature and other elements like frustration interaction could be added to induce the radical destruction of Kondo screening in lattice fermion models.
We expect that the findings found here may be helpful for deeper understanding on interacting topological Kondo insulator candidate SmB$_{6}$ and other related quantum topological materials.

The remainder of this paper is organized as follows: In Sec.~\ref{sec2}, the model is introduced and its non-interacting feature is analyzed in detail.
In Sec.~\ref{sec3}, PQMC is performed and a Haldane-like phase is found by inspecting the site-resolved magnetization and strangle correlator. Sec.~\ref{sec4} gives further information on
edge magnetization under different interaction and hybridization strength. Discussions on the origin of magnetic end state, surface Kondo breakdown, spin-only model and possible optical lattice realization are given in Sec.~\ref{sec5}.
Finally, conclusions are presented in Sec.~\ref{sec6} with provision of a possible direction for future work.

\section{$1D$ $p$-wave periodic Anderson model}\label{sec2}
The model we have studied is the following $1D$ $p$-wave periodic Anderson model:\cite{Dzero2010,Mezio2015,Coleman2015}
\begin{eqnarray}
H&&=\sum_{j\sigma}[t_{c}c_{j\sigma}^{\dag}c_{j+1\sigma}-t_{f}f_{j\sigma}^{\dag}f_{j+1\sigma}+\mathrm{H.c.}]\nonumber \\
&&+\frac{V}{2}\sum_{j\sigma}[(c_{j+1\sigma}^{\dag}-c_{j-1\sigma}^{\dag})f_{j\sigma}+f_{j\sigma}^{\dag}(c_{j+1\sigma}-c_{j-1\sigma})]\nonumber \\
&&+E_{f}\sum_{j\sigma}f_{j\sigma}^{\dag}f_{j\sigma}+U\sum_{j}f_{j\uparrow}^{\dag}f_{j\uparrow}f_{j\downarrow}^{\dag}f_{j\downarrow}\label{eq1}.
\end{eqnarray}
Here, $t_{c}$ and $t_{f}$ are the nearest-neighbor-hopping strengths, $E_{f}$ denotes the energy level of f-electron and local electron has also the conventional Hubbard on-site interaction ($U$-term).
The $p$-wave hybridization between conduction and local electron is encoded by $V$ term, in which the coupling of conduction and local electron is non-local (due to non-trivial spin-orbit coupling between $d$-like conduction electron and $f$-like local electron) in contrast to the usual
$s$-wave (on-site) hybridization in standard periodic Anderson model. (See also Fig.~\ref{fig:PAM})
As we will see below, such non-local hybridization leads to non-trivial topological phases and the main object of this article is to discuss their detailed features.

As we are only interested in insulating topological states in present work, the system is fixed to be half-filled, which means the total number of conduction and local electrons is equal to twice of lattice site number. Next, the hopping parameters $t_{c}$ and $t_{f}$ should have the same sign,\cite{Erten2016,Legner2014} i.e. $t_{c}t_{f}>0$, otherwise
even the non-interacting energy band will not be fully gapped and the topological argument will be meaningless. Furthermore, we choose $E_{f}=-U/2$, such that
when interaction is turned on, it will exclude the fermion minus-sign problem for PQMC, which will be used to explore the effect of interaction.\cite{Assaad2008}

In addition, in terms of a canonical transformation (i.e., a generalized Schrieffer-Wolff transformation), the model in Eq.~\ref{eq1} can lead to the $1D$ $p$-wave Kondo lattice model which was firstly proposed by Alexandrov and Coleman, and then
further studied in Refs.~\cite{Alexandrov2014,Lobos2015,Mezio2015,Hagymasi2016}. (One finds $J_{K}=\frac{4t_{f}^{2}}{U}$ and $J_{H}=\frac{2V^{2}}{U}$ with our notation for $V$.)

\subsection{Non-interacting limit: Bulk property}
Firstly, we discuss an exactly soluble limit, where the Hubbard interaction is turned off ($U=0$),
\begin{eqnarray}
H_{0}=\sum_{k}\psi_{k}^{\dag}
\left(
  \begin{array}{cc}
    \varepsilon_{c}(k)\hat{I} & iVs_{x}\hat{\sigma}_{x} \\
    -iVs_{x}\hat{\sigma}_{x} & \varepsilon_{f}(k)\hat{I} \\
  \end{array}
\right)
\psi_{k}=\sum_{k}\psi_{k}^{\dag}\mathcal{H}(\vec{k})\psi_{k}\nonumber.
\end{eqnarray}
Here, the model has been transformed into momentum space via Fourier transformation and a four-component spinor $\psi_{k}$ has been introduced as $\psi_{k}=(c_{k\uparrow}, c_{k\downarrow}, f_{k\downarrow}, f_{k\uparrow} )^{T}$. $\hat{\sigma}_{x},\hat{\sigma}_{y},\hat{\sigma}_{z}$ are the usual $2\times2$ Pauli matrices acting on orbital basis and $\hat{I}$ is the $2\times2$ unit matrix.
The dispersion of conduction and f-electron is given by
$\varepsilon_{c}(k)=2t_{c}\cos k_{x}$
and $\varepsilon_{f}(k)=-2t_{f}\cos k_{x}$, respectively. Moreover, the $p$-wave hybridization is denoted by $Vs_{x}\hat{\sigma}_{x}$ with form factor $s_{x}=\sin k_{x}$.

For this non-interacting model, its quasi-particle energy spectrum is easily found to be
\begin{eqnarray}
E_{k\pm}&&=\frac{\varepsilon_{c}(k)+\varepsilon_{f}(k)\pm\sqrt{(\varepsilon_{c}(k)-\varepsilon_{f}(k))^{2}+4V^{2}s_{x}^{2}}}{2}\nonumber\\
&&=(t_{c}-t_{f})\cos k_{x}\pm\sqrt{(t_{c}+t_{f})^{2}\cos^{2}k_{x}+V^{2}\sin^{2}k_{x}}\nonumber\\
\label{eq2},
\end{eqnarray}
which has a two-fold spin degeneracy. When the system is half-filled, the lower band $E_{k-}$ is fully occupied and a gap exists for all
quasi-particle excitations, thus we are truly considering a bulk insulator.

Following Ref.~\cite{Dzero2010}, the system $H_{0}$ is invariant under both time-reversal $\mathcal{T}$ and space-inversion $\mathcal{P}$ transformation
since $\mathcal{H}(\vec{k})^{T}=\mathcal{T}\mathcal{H}(-\vec{k})\mathcal{T}^{-1}$
and $\mathcal{H}(\vec{k})=\mathcal{P}\mathcal{H}(-\vec{k})\mathcal{P}^{-1}$
with
\begin{equation}
\mathcal{T}=\left(
  \begin{array}{cc}
    i\hat{\sigma}_{y} & 0 \\
    0 & i\hat{\sigma}_{y} \\
  \end{array}
\right),~~~\mathcal{P}=\left(
  \begin{array}{cc}
    \hat{I} & 0 \\
    0 & -\hat{I} \\
  \end{array}
\right)\nonumber.
\end{equation}
So, the non-interacting Hamiltonian $H_{0}$ has time-reversal and inversion symmetry, and according to the Fu-Kane formula,\cite{Fu2007}
the system can be a $Z_{2}$ topological insulator if the following $Z_{2}$ index $\nu=1$:
\begin{equation}
(-1)^{\nu}=\delta_{\Gamma}\delta_{M},\nonumber
\end{equation}
Here, $\delta_{\Gamma},\delta_{M}$ are the parity at the high-symmetry points of $1D$ Brillouin zone. ($\Gamma=0$ and $M=\pi$)

To find the parity at those two high-symmetry points, we rewrite $\mathcal{H}(\vec{k})$ as
\begin{eqnarray}
\mathcal{H}(\vec{k})&&=(\varepsilon_{c}(k)+\varepsilon_{f}(k))\left(
  \begin{array}{cc}
    \hat{I} & 0 \\
    0 & \hat{I} \\
  \end{array}
\right)+(\varepsilon_{c}(k)-\varepsilon_{f}(k))\left(
  \begin{array}{cc}
    \hat{I} & 0 \\
    0 & -\hat{I} \\
  \end{array}
\right)\nonumber\\
&&+iVs_{x}\left(
  \begin{array}{cc}
    0 & \hat{\sigma}_{x} \\
    -\hat{\sigma}_{x} & 0 \\
  \end{array}
\right)\nonumber.
\end{eqnarray}
Then, at high-symmetry points $\Gamma$ and $M$, the hybridization term vanishes since $s_{x}=\sin 0=\sin \pi=0$
, which is a general result from time-reversal symmetry. Therefore, the parity is determined by relative position of $\varepsilon_{c}(k)$ and
$\varepsilon_{f}(k)$, i.e.
\begin{eqnarray}
&&\delta_{\Gamma}=\mathrm{sign}(\varepsilon_{c}(0)-\varepsilon_{f}(0))=\mathrm{sign}[t_{c}+t_{f}]\nonumber\\
&&\delta_{M}=\mathrm{sign}(\varepsilon_{c}(\pi)-\varepsilon_{f}(\pi))=-\mathrm{sign}[t_{c}+t_{f}]\nonumber.
\end{eqnarray}
So, we find $\nu=1$ ($(-1)^{\nu}=\delta_{\Gamma}\delta_{M}=-1$) and the half-filled system is in fact a $1D$ $Z_{2}$ topological insulator.
(In fact, if the f-electron level $E_{f}$ is reintroduced,
we find the non-interacting system is still a $Z_{2}$ topological insulator when $-2|t_{c}+t_{f}|<E_{f}<2|t_{c}+t_{f}|$, given the half-filling condition is satisfied and chemical potential is properly tuned.)
\subsection{Non-interacting limit: Edge property}
Alternatively, one may explore the topological property by inspecting the edge states of Hamiltonian $H_{0}$ in real space with open boundary condition:
\begin{eqnarray}
H_{0}&&=\sum_{j=1,\sigma}^{L-1}[t_{c}c_{j\sigma}^{\dag}c_{j+1\sigma}-t_{f}f_{j\sigma}^{\dag}f_{j+1\sigma}+\mathrm{H.c.}]\nonumber \\
&&+\frac{V}{2}\sum_{j=2,\sigma}^{L-1}[(c_{j+1\sigma}^{\dag}-c_{j-1\sigma}^{\dag})f_{j\sigma}+f_{j\sigma}^{\dag}(c_{j+1\sigma}-c_{j-1\sigma})]\nonumber\\
&&+\frac{V}{2}\sum_{\sigma}[c_{2\sigma}^{\dag}f_{1\sigma}+f_{1\sigma}^{\dag}c_{2\sigma}+c_{L-1,\sigma}^{\dag}f_{L\sigma}+f_{L\sigma}^{\dag}c_{L-1,\sigma}]\nonumber\\
\label{eq3}.
\end{eqnarray}

To elucidate the edge state, we consider a special case with $t_{c}=t_{f}=t$, such that
\begin{eqnarray}
H_{0}(t)&&=(t+\frac{V}{2})\sum_{j=1,\sigma}^{L-1}[b_{j+1\sigma}^{\dag}a_{j\sigma}+a_{j\sigma}^{\dag}b_{j+1\sigma}]\nonumber\\
&&+(t-\frac{V}{2})\sum_{j=2,\sigma}^{L}[b_{j-1\sigma}^{\dag}a_{j\sigma}+a_{j\sigma}^{\dag}b_{j-1\sigma}]\nonumber.
\end{eqnarray}
We introduce new (bonding and anti-bonding) fermions $a_{j\sigma}=\frac{1}{\sqrt{2}}(f_{j\sigma}-c_{j\sigma})$, and $b_{j\sigma}=\frac{1}{\sqrt{2}}(f_{j\sigma}+c_{j\sigma})$.
Furthermore, if $t=V/2$ (this is the so-called 'Kitaev point' and $t=-V/2$ case is similar), we have
\begin{eqnarray}
H_{0}(t=V/2)=2t\sum_{j=2,\sigma}^{L}[b_{j-1\sigma}^{\dag}a_{j\sigma}+a_{j\sigma}^{\dag}b_{j-1\sigma}]\nonumber,
\end{eqnarray}
which means zero energy fermion modes created by $a_{1\sigma}^{\dag}$ and $b_{L\sigma}^{\dag}$ are decoupled from Hamiltonian $H_{0}(t=V/2)$.
In other words, the expected edge states are actually formed by these fermion zero modes like $|\Psi_{edge}\rangle\sim a_{1\uparrow}^{\dag}a_{1\downarrow}^{\dag}|0\rangle$,
$a_{1\uparrow}^{\dag}b_{L\downarrow}^{\dag}|0\rangle$, $a_{1\downarrow}^{\dag}b_{L\uparrow}^{\dag}|0\rangle$, and $b_{L\uparrow}^{\dag}b_{L\downarrow}^{\dag}|0\rangle$, where one is able to obtain four-fold degenerated many-body ground-state wave-function for this non-interacting case.

\begin{figure}
\resizebox{0.5\textwidth}{!}{%
  \includegraphics{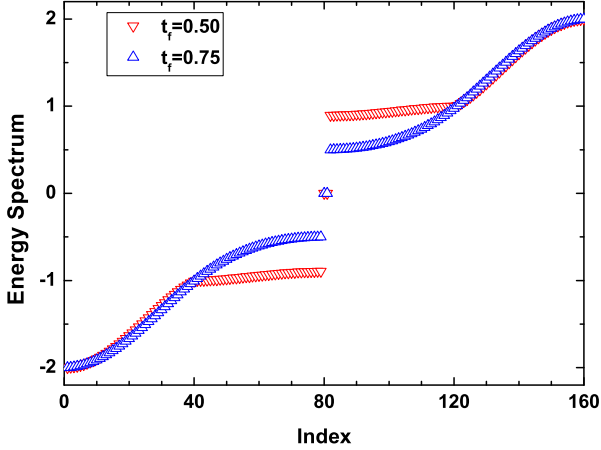}%
}
\caption{Fermion zero mode (edge state) per spin-flavor in non-interacting $1D$ p-wave periodic Anderson model. Parameters are set to $L=80$, $t_{c}=1$ and $V=1$.}
\label{fig:TKI_0}
\end{figure}

More generally, when $t_{c}\neq t_{f}\neq V/2$, these edge modes are still stable if no bulk gap is closed during evolution of Hamiltonian.
(e.g. see Fig.\ref{fig:TKI_0})
In addition, we have checked that for all the cases with $t_{c}t_{f}>0$ and $V\neq0$, the system shows fermion zero modes, which means the system itself is always in a topological state.
This finding is consistent with the results of topological band theory (calculation of $Z_{2}$ index $\nu=1$) seen in last subsection.

\section{Turning on interaction: Brute-force quantum Monte Carlo simulation}\label{sec3}
From the analysis presented in the last section, we know that the
non-interacting system is located in a $Z_{2}$ topological insulating phase with fermion edge mode.

When we turn on interaction $U$, a complicated situation
may occur. For example, if adiabatic continuity works,
then we expect the $U =0$ topological phase to remain the same
for not too large $U$. In contrast, particularly for $1D$
system, the effect of interaction is radical and even a small
$U$ can lead the system into a new phase different from
the non-interacting limit.

Here, we will use PQMC to simulate the $1D$ $p$-wave periodic Anderson model.
Basically, in PQMC, one is able to calculate ground-state expectation value of observable $\mathcal{O}$ as
\begin{equation}
\langle \mathcal{O}\rangle=\frac{\langle\Psi_{g}|\mathcal{O}|\Psi_{g}\rangle}{\langle\Psi_{g}|\Psi_{g}\rangle}=\lim_{\beta\rightarrow\infty}\frac{\langle\Psi_{T}|e^{-\frac{\beta}{2} \hat{H}}\mathcal{O}e^{-\frac{\beta}{2} \hat{H}}|\Psi_{T}\rangle}{\langle\Psi_{T}|e^{-\beta \hat{H}}|\Psi_{T}\rangle}    \nonumber
\end{equation}
where $|\Psi_{g}\rangle$ is the ground-state many-body wave-function of Eq.~\ref{eq1}, obtained from a imaginary-time projection of trial wave-function $|\Psi_{T}\rangle$.
This trial wave-function can be simply chosen as the ground-state of the non-interacting Hamiltonian, e.g. $H_{0}$ (Eq.~\ref{eq3}), or a Hartree-Fock mean-field solution of the whole Hamiltonian.
Moreover, the requirement $\langle\Psi_{g}|\Psi_{T}\rangle\neq0$ should be fulfilled, otherwise the above projection method will not converge into the desirable interacting many-body ground-state.
In realistic numerical calculation, the imaginary-time length $\beta$ is obviously finite but a large value of it is able to obtain a convergent result. (e.g. $\beta>L$ with $L$ being the size of the system)

Motivated by the previous study of $1D$ $p$-wave Kondo lattice, we will study this model by fixing hopping energy $t_{c}=t_{f}=1$ such that
a finite $U$ case may lead to visible Haldane state with insulating bulk and free local moments located on the boundary.
Magnetization on each site and string order parameter may be used in order to find such phase however for the PQMC, the calculation of exponential of operators in string order
is very challenging, thus we may calculate site-resolved magnetization to detect the possible edge local moment.

\subsection{Benchmark: Non-interacting limit}
\begin{figure}
\resizebox{0.5\textwidth}{!}{%
  \includegraphics{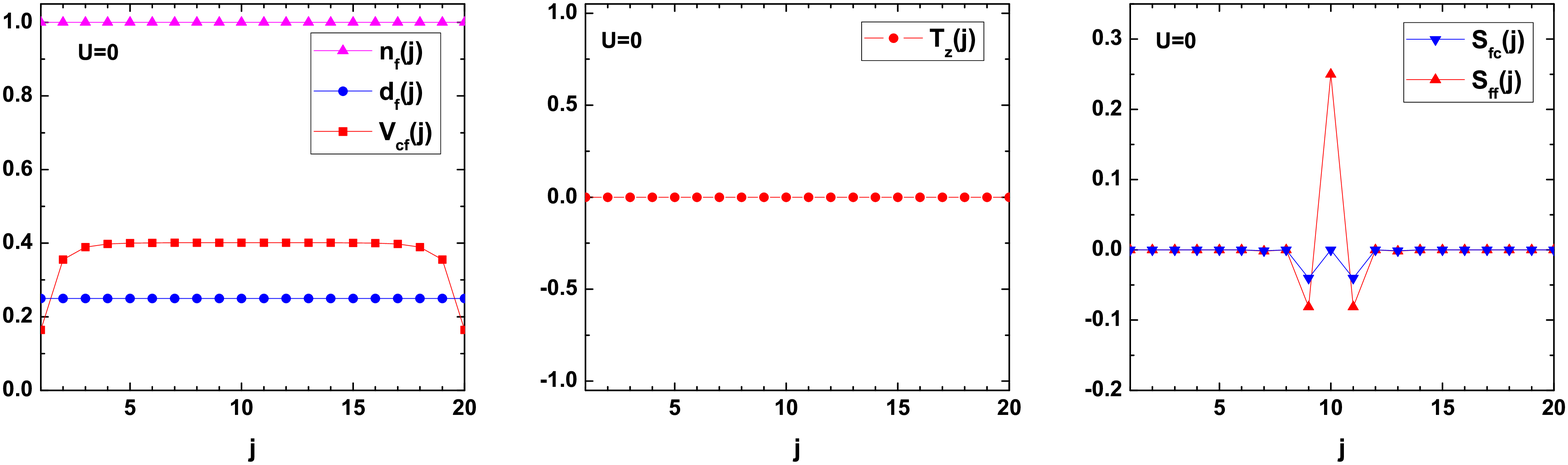}%
}
\caption{Non-interacting limit of $1D$ $p$-wave periodic Anderson model. Parameters are $L=20$, $t_{c}=t_{f}=1$, $V=1$ and $U=0$. (Left) f-electron density $n_{f}(j)$,
double occupation number of f-electron $d_{f}(j)$, and $c-f$ hybridization $V_{cf}(j)$; (middle) site-resolved magnetization $T_{z}(j)$; (right) spin correlations function between f-electron and conduction electron  $S_{fc}(j)$, and f-electron $S_{ff}(j)$.}
\label{fig:1DPAM_U=0}
\end{figure}

Before discussing the interaction problem, we first show results in non-interacting limit ($U=0$) with open boundary condition.
From Fig.~\ref{fig:1DPAM_U=0}, we can see that the f-electron density $n_{f}(j)=\sum_{\sigma}\langle f_{j\sigma}^{\dag}f_{j\sigma}\rangle$ is uniform and the site-resolved magnetization $T_{z}(j)=\langle S_{z}^{f}(j)\rangle+\langle S_{z}^{c}(j)\rangle$ is zero for all sites.
The double occupation number of f-electron $d_{f}(j)=\langle n_{\uparrow}^{f}n_{\downarrow}^{f}\rangle=\langle f_{\uparrow}^{\dag}f_{\uparrow}f_{\downarrow}^{\dag}f_{\downarrow}\rangle$ is
at its non-interacting limit, which is $d_{f}(j)=\langle n_{\uparrow}^{f}\rangle\langle n_{\downarrow}^{f}\rangle=(1/2)^{2}=0.25$ for single occupation on each site. The
$c-f$ hybridization $V_{cf}(j)=-\frac{1}{2}\sum_{\sigma}\langle f_{j\sigma}^{\dag}(c_{j+1\sigma}-c_{j-1\sigma})\rangle$ is found to be weakened at boundary, where fewer conduction electron nearby reduces
the hybridization. Moreover, the spin correlations $S_{fc}(j)=\langle S_{z}^{f}(L/2)S_{z}^{c}(j)\rangle$ and $S_{ff}(j)=\langle S_{z}^{f}(L/2)S_{z}^{f}(j)\rangle$ are
all short-ranged. (The typical correlation length in this case is about one or two sites.) So, the system should be in a spin-disordered state while the $V=0$ case with
a much longer correlation length ($5$ to $10$ sites) is a metallic state. The charge correlation (not shown here) is similar.

When we consider bulk system (with periodic boundary condition), the spin, charge and single-particle gap above the half-filled ground-state can be estimated as
\begin{eqnarray}
&&\Delta_{s}=E(L+1,L-1)-E(L,L)\nonumber\\
&&\Delta_{c}=E(L+1,L+1)-E(L,L)\nonumber\\
&&\Delta_{sp}=2(E(L+1,L)-E(L,L))\nonumber
\end{eqnarray}
where $E(N_{\uparrow},N_{\downarrow})$ denotes the ground-state energy with $N_{\uparrow}$ referring to  spin-up electron and $N_{\downarrow}$ spin-down electron.
For $L=20$, $t_{c}=t_{f}=V=1$, it is found that $\Delta_{s}=\Delta_{c}=\Delta_{sp}=2$, which is consistent with direct calculation using Eq.~\ref{eq2}. So, the bulk system is an insulator with both finite charge and spin gap.
Now, we consider the open boundary system, where we obtain $\Delta_{s}=\Delta_{c}=\Delta_{sp}=0$, which means zero-energy (fermion edge) mode must exist and it agrees with the finding from the last
section.

Therefore, we conclude that the $U=0$ system is in a bulk insulating state with both charge and spin gap, and its edge has a fermion zero-energy mode.
An important indication from here is that there is no free spin-$1/2$ local moment at edge and we conclude that the $U=0$ state is obviously not the Haldane phase but the $Z_{2}$ topological insulator discussed previously.
\subsection{Interacting case and possible Haldane-like phase}
\begin{figure}
\resizebox{0.5\textwidth}{!}{%
  \includegraphics{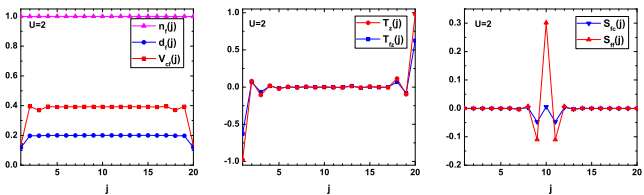}%
}
\caption{Interacting effect on $1D$ $p$-wave periodic Anderson model. Parameters are $L=20$, $t_{c}=t_{f}=1$, $V=1$ and $U=2$.
(Left) f-electron density $n_{f}(j)$, double occupation number of f-electron $d_{f}(j)$, and $c-f$ hybridization $V_{cf}(j)$; (middle) site-resolved magnetization $T_{z}(j)$ and its f-electron component $T_{fz}(j)$; (right) spin correlations function between f-electron and conduction electron $S_{fc}(j)$, and f-electron $S_{ff}(j)$.}
\label{fig:1DPAM_U=2}
\end{figure}
For interaction effect, we now study a $20$-site chain
with $U = 2$, $V = 1$ and $\beta=30$. We have also tested
a longer chain with $L = 32$ and it does not lead to any
sensible changes compared to $L = 20$ chain.
Additionally, statistical errors in our PQMC simulation are typically smaller than the symbol size in the plot and will not be explicitly shown in figures.

From Fig.\ref{fig:1DPAM_U=2}, we see that the double occupation number is suppressed around the boundary, which is an effect of interaction since the non-interacting case shows uniform distribution.
Such reduced double occupation possibility actually leads to an enhanced local moment in edge site. The effective $c-f$ hybridization is reduced around edge like the non-interacting case.
From these two results, we may assume the local moment at boundary in the interacting situation is more free than its non-interacting counterpart.

The most interesting result can be drawn from the site-resolved magnetization $T_{z}(j)$, where the large magnetization (equalling to nearly a unit) is observed at two edges.\cite{Mezio2015}
This may be seen as a free $1/2$-local moment/spin localized at boundary while the inner sites only have small magnetization, which reflects the nature of bounded spin-singlet.
So, this indicates there exists so-called magnetic end state proposed in large-N mean-field calculation and found in bosonization and DMRG.\cite{Alexandrov2014,Lobos2015,Mezio2015,Hagymasi2016}
However, this end state is not corresponding to the non-interacting limit since the latter has no magnetization around the edge.
For comparison, the site-resolved magnetization of local electron $T_{fz}(j)$ is also shown and it implies that the total magnetization
at boundary is most contributed from the local $f$-electron and the remaining conduction electron has smaller contribution.

In addition, the short-ranged spin-spin correlation for conduction and f-electron show ferromagnetic correlation for on-site spins, while the nearest ones are anti-ferromagnetic.
The same case is found for spin correlation between f-electrons. Such that, if we approximate the system as two coupled spin-$1/2$ chains, then
the effective coupling between chains is ferromagnetic. Therefore, in the low-energy limit, a Haldane-like phase should form and the corresponding edge
state has a free spin-$1/2$ magnetic moment, which agrees with the observation in the magnetization.

Now, we conclude that the interacting model shows the expected Haldane-like phase with free (spin-$1/2$) magnetic moment situated at the boundary, consistent with DMRG and bosonization,\cite{Lobos2015,Mezio2015,Hagymasi2016} but
in contrast to the non-interacting case. So, we have seen that the adiabatic continuity breaks down in this model when interaction is added and
the interaction effect is fundamental in $1D$, where sophisticated techniques beyond mean-field theories should be used.

\subsection{Strange correlator in $1D$ p-wave periodic Anderson model}
Recently it has been proposed that the strange correlator is able to detect short-range entangled topological states and
has been successful in identifying many topological states of matter including
the Haldane phase in spin-one Heisenberg antiferromagnets, $1D$ and $2D$ AKLT states, quantum spin Hall state
and bosonic symmetry-protected topological state.\cite{You2014,Wierschem2014,Wu2015,He2016}
The virtue of strange correlator is that no bipartition of a system is involved and
the finite-size effect from the open boundary calculation is heavily reduced.

For our cases, we use PQMC to sample the target ground-state wave-function from the imaginary evolution of a trial wave-function $|\Psi_{0}\rangle$ while
a trivial state $|\Omega\rangle$ is chosen to be a state with $E_{f}=10$. Thus, the following strangle correlator calculated
\begin{figure}
\resizebox{0.5\textwidth}{!}{%
  \includegraphics{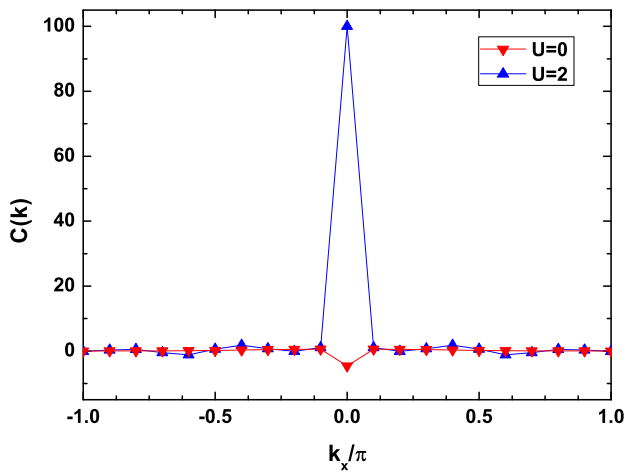}%
}
\caption{Strange correlator $C(k)$ in $1D$ p-wave periodic Anderson model, $U=0$ versus $U=2$. The divergent point is at $k_{x}=0$ while the finite-size cutoff is also seen.}
\label{fig:PQMC_strange}
\end{figure}
\begin{equation}
C(k)=\frac{1}{L^{2}}\sum_{j,l}e^{i(j-l)k}C_{jl}=\frac{1}{L^{2}}\sum_{j,l}\frac{\langle\Omega|c_{j\sigma}c_{l\sigma}^{\dag}|\Psi\rangle}{\langle\Omega|\Psi\rangle}\nonumber
\end{equation}
In Fig.~\ref{fig:PQMC_strange}, we see that both the non-interacting case ($U=0$) and interacting case ($U=2$) show
divergent-like behavior in their strange correlators, which means they are all non-trivial topological states,
although the divergence is cut off by finite-size effect.
Interestingly, the interacting system has stronger divergence than the non-interacting one, which may indicate that
the Haldane-like phase may be more entangled than its non-interacting counterpart, the $Z_{2}$ topological insulating state.
The results here supplement the observation in site-resolved magnetization and agree with each other.

\section{Edge magnetization under different interaction $U$ and hybridization $V$}\label{sec4}
In the last section, we have seen that a Haldane-like phase is found in our $1D$ $p$-wave periodic Anderson model,
whose essential feature is the existence of free magnetic moment at the boundary of an open chain system and is detected by non-vanished
magnetization of both conduction and local electron.

In this section, we proceed to see the evolution of edge magnetization under different interaction strength $U$ and hybridization $V$.
This may encode how edge magnetic moment is free from the ones in a bulk system.
\begin{figure}
\resizebox{0.5\textwidth}{!}{%
  \includegraphics{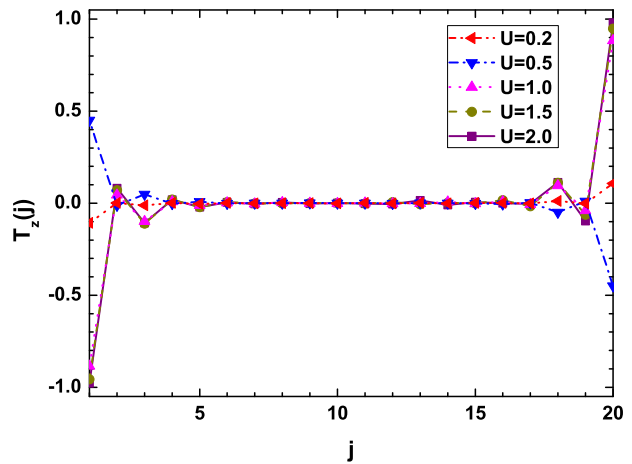}%
}
\caption{Edge magnetization versus different Hubbard interaction $U$. Other parameters are $L=20$, $t_{c}=t_{f}=1$ and $V=1$.}
\label{fig:PQMC_U_change}
\end{figure}

\begin{figure}
\resizebox{0.5\textwidth}{!}{%
  \includegraphics{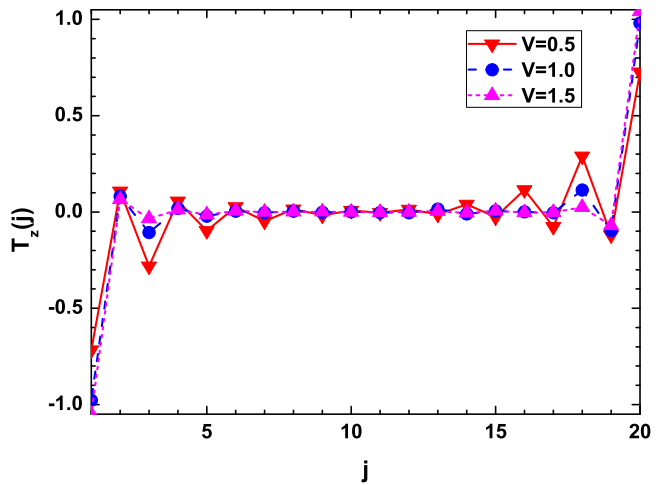}%
}
\caption{Edge magnetization versus different hybridization strength $V$. Other parameters are $L=20$, $t_{c}=t_{f}=1$ and $U=2$.}
\label{fig:PQMC_V_change}
\end{figure}
\subsection{Effect of Hubbard interaction $U$ on the edge magnetization}
First, we change the strength of Hubbbard interaction $U$ while keeping other parameters intact. ($t_{c}=t_{f}=V=1$) In Fig.~\ref{fig:PQMC_U_change}, we find that
the edge magnetization increases from zero to unit when interaction $U$ is gradually enhanced. (When $U=0$, magnetization is vanished as seen in Fig.~\ref{fig:1DPAM_U=0}.)
It is noted that the case with $U=0.5$ has inverted magnetization in comparison to other cases and
the reason is that in all numerical simulations performed in this work, we consider a total spin-singlet system
but since edge magnetization is able to build at two equivalent direction (e.g. spin-up and spin down direction), it inevitably leads to two degenerated ground-states.
However, for PQMC, it only randomly selects one of ground-states and this explains the observed distinction.

\subsection{Effect of hybridization strength $V$}
Next, tuning different values of hybridization strength $V$ leads to the results shown in Fig.~\ref{fig:PQMC_V_change}.
Here, we have observed that
larger $V$ results in the smaller edge magnetization, which is due to the enhanced Kondo screening and charge fluctuation for larger $V$. Specifically, the system with $V=1.5$ has boundary magnetization $T_{z}\approx0.7$ compared to $T_{z}\approx1.0$ for $V=1.0$ case. Furthermore, when $V$ is as small as $0.5$ for fixed other parameters, magnetization at boundary sites (see red dots for $j=1$ and $j=20$ in Fig.~\ref{fig:PQMC_V_change}) actually exceeds unit.

\subsection{Why edge magnetization is not unit}
After all, we find that these results show that with generic parameters chosen for our model, the edge magnetization is deviated from unit
as in pure spin-one antiferromagnetic chain, where decoupled (fractionalized) spin-half objects appear at the boundary.

In our present model, charge fluctuation of local electron always exists due to the hybridization $V$ between local and conduction electron unlike the previously studied p-wave Kondo lattice model, where
charge fluctuation is completely excluded by canonical transformation with assumption of $V/U<<1$. For a strong hybridization (e.g. $V\gg U$), empty, spin-up, spin-down and double occupation are all allowed and it is not reasonable to expect a freezed unit magnetization at boundary when its particle occupation is rapidly changed among these four states. Instead, in this condition, due to the mentioned charge fluctuation, we expect a reduced magnetization at edge sites.
In the opposite limit, $V$ is smaller than $U$ and Hubbard interaction is able to stabilize
local moment, thus a large magnetization at boundary is possible in this case. Moreover, as has been discussed in Sec.3.2, we may think that the on-site conduction and local electron have ferromagnetic coupling, thus the total edge magnetization may exceed unit since two kinds of electron can contribute.

In addition, we may imagine that a putative spin-$1/2$ object at boundary will fluctuate into a spin-one composite or a zero-spin one, thus the effective magnetic moment
at edge is not unit but changed by its surrounding electron configuration.
It is noted that similar results have been found in doped three-legged Hubbard ladder, where the insulating phase is always the Haldane phase regardless
of the suppression from unity of spin.\cite{Nourse2016}

\section{Discussions}\label{sec5}

\subsection{Why magnetic end states appear in interacting system}
Here, we provide a qualitative argument on why magnetic end states appear in interacting system.
Firstly, we know that when $t_{c}=t_{f}=V/2$ (Kitaev point), the non-interacting ground-state under open boundary condition can be approximated as
a superposition of four zero-modes $a_{1\uparrow}^{\dag}a_{1\downarrow}^{\dag}|0\rangle,
a_{1\uparrow}^{\dag}b_{L\downarrow}^{\dag}|0\rangle,a_{1\downarrow}^{\dag}b_{L\uparrow}^{\dag}|0\rangle,b_{L\uparrow}^{\dag}b_{L\downarrow}^{\dag}|0\rangle$.
In this case, the contribution of these four degenerated states is equal and the magnetization is obviously zero and no magnetic end state exists.

However, when Hubbard-$U$ interaction is turned on, the double-occupation on local electron site is suppressed, thus states like $a_{1\uparrow}^{\dag}a_{1\downarrow}^{\dag}|0\rangle$ and
$b_{L\uparrow}^{\dag}b_{L\downarrow}^{\dag}|0\rangle$ have small contribution to the system and the remaining ones are $a_{1\uparrow}^{\dag}b_{L\downarrow}^{\dag}|0\rangle$ and
$a_{1\downarrow}^{\dag}b_{L\uparrow}^{\dag}|0\rangle$. Clearly, these two survivals are degenerated, but for numerical simulation like PQMC performed in this work,
one is only able to obtain one of these two states at a time. In other words, in one simulation, we may obtain $a_{1\uparrow}^{\dag}b_{L\downarrow}^{\dag}|0\rangle$ as our ground-state while in other
simulation, $a_{1\downarrow}^{\dag}b_{L\uparrow}^{\dag}|0\rangle$ may be observed. (See e.g. Fig.~\ref{fig:PQMC_U_change})
Therefore, for each of state, we can observe the magnetic end mode. More generally, since the topological feature is not changed under (smooth) continuous mapping of Hamiltonian,
we expect the argument at special point $t_{c}=t_{f}=V/2$ is still valid at least qualitatively for generic conditions.
In some sense, the interaction in our $p$-wave periodic Anderson model reduces the four-fold degeneracy of non-interacting ground-state into double degeneracy for interacting ground-state, which is similar to the findings
in Su-Schrieffer-Heeger-Hubbard model and symmetry-protected one-dimensional
fermionic superconducting phases.\cite{Wang2015,Tang2012}

\subsection{How about surface Kondo breakdown}
In Refs.~\cite{Alexandrov2015,Thomson2016,Erten2016,Alexandrov2014}, authors have proposed that for
topological Kondo insulator, whose $c-f$ hybridization is non-local in nature, Kondo screening at boundary may be weakened or even not developed
due to smaller number of nearly electrons. In Sec.~\ref{sec3}, we have seen that the effective $c-f$ hybridization $V_{cf}$ is indeed
weakened at boundary sites but its value is larger than $0.1$. (recall $V=1$ in these cases) Thus, no breakdown of Kondo screening is found in this uniform $V$ case.

A careful reader may note that a more conventional quantity to measure the strength of Kondo screening
is spin-singlet correlator, i.e. c-f spin correlation function like $\langle\vec{S}^{f}(i)\cdot\vec{S}^{c}(j)\rangle$. (see e.g. Ref.~\cite{Paiva2003}) We have checked that the results of our simplified definition of $V_{cf}$ are physically consistent with the more conventional spin-singlet correlator.

\begin{figure}
\resizebox{0.5\textwidth}{!}{%
  \includegraphics{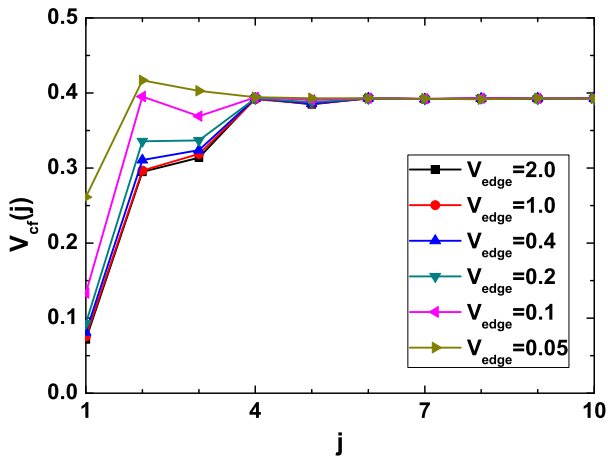}%
}
\caption{Effective $c-f$ hybridization $V_{cf}(j)$ as a function of site index $j\in[1,L/2]$. Different lines correspond to different edge hybridization $V_{edge}$, respectively. Other parameters are $L=20$, $t_{c}=t_{f}=V=1$ and $U=2$.}
\label{fig:Vcf_edge}
\end{figure}

Now, following the suggestion in Refs.~\cite{Alexandrov2015,Alexandrov2014},
in order to see possible Kondo breakdown at boundary, one may decrease the bare $c-f$ hybridization $V_{edge}$ at edge site while keeping bulk $V$ intact.
In Fig.\ref{fig:Vcf_edge}, we have chosen $V_{edge}=2,1,0.4$,
$0.2,0.1,0.05$ (bulk $V$ is fixed to unit) and it is found that the $c-f$ hybridization $V_{cf}$
is indeed reduced but still finite even though the edge hybridization $V_{edge}$ is much smaller than the bulk hybridization $V$. Therefore,
no surface Kondo breakdown is found in our numerical simulation at zero temperature.

However, it is important to emphasize that we do not conclude that the surface Kondo breakdown would not take place in realistic 3D materials like SmB$_{6}$. Our study here is just a simplified 1D lattice model, and is obviously not invented for the real-life materials. The only thing we have obtained here is that at least for this simplified 1D TKI model at zero $T$, the Kondo breakdown is not observed in our numerical calculation. Although we do not find clue of Kondo breakdown in our present work,
we expect a well devised 3D TKI model (but beyond our present numerical technique because both 2D and 3D TKI models in literature have severe fermion minus-sign problem) may support Kondo breakdown even at $T=0$.
\subsection{Spin-only model}
In the main text, we have studied the $p$-wave periodic Anderson model, where both charge and spin degree of freedom of electrons are included.
Here, it is interesting to think whether a spin-only model may capture the topological phases found in Kondo lattice and periodic Anderson-like models.
In literature, $1D$ Kondo necklace model might be the desirable one and we propose its $p$-wave version as follows\cite{Doniach1977,Zhong2013}
\begin{eqnarray}
H_{p-KN}&&=-t\sum_{j}(\tau_{j}^{x}\tau_{j+1}^{x}+\tau_{j}^{y}\tau_{j+1}^{y})\nonumber\\
&&+J\sum_{j}(\vec{\tau}_{j+1}-\vec{\tau}_{j-1})\cdot\vec{S}_{j}\nonumber
\end{eqnarray}
where $\tau_{i}^{x},\tau_{i}^{y}$ represents the spin degrees of freedom coming from original conduction electron and $\vec{S}_{i}$ denotes local moment of f-electron.
The $J$-term denotes the $p$-wave-like 'Kondo' coupling and it has staggered features,
which may lead to effective ferromagnetic coupling between these two chains and we expect a Haldane-like phase can form in this model.\cite{Zhong2014} However, we should remind the reader that the Kondo necklace model cannot be derived from the original Kondo lattice model at half-filling but can only be considered as a phenomenological model devised for studying the low-lying spin excitations.

\subsection{Realization in cold-atom setup}
As proposed in Ref.~\cite{Lobos2015}, the $p$-wave periodic Anderson model may be realized by $p$-band optical
lattices with a balanced mixture of two-component Fermi atoms (e.g. $^{6}$Li or $^{40}$K).
On the other hand, one may use ultracold alkaline earth-like atoms, e.g. $^{173}$Yb, to simulate our calculated model via suitable laser excitations.\cite{Nakagawa2015,Zhong2016}
However, it should be emphasized that the reachable temperature in current optical lattice experiments is still too high to observe the low-temperature
Kondo (lattice) physics and the corresponding topological phases. Thus, preliminary exploration on finite temperature regime for our model will be more
useful for future experiments on this interesting topic.

\section{Conclusion and direction for future work}\label{sec6}
In conclusion, we have studied the $p$-wave periodic Anderson model (Eq.~\ref{eq1}) in terms of the unbiased PQMC simulation.
The ground-state is the expected Haldane-like phase with magnetic end mode driven by electron correlation effect beyond effective non-interacting single-particle picture.
In addition, we have not found any evidence of the surface Kondo breakdown proposed in the literature at least for
zero temperature, so it is suspected that frustration-like interaction, e.g. ring-exchange interaction among nearest sites, may be crucial in inducing such radical destruction of Kondo screening in the lattice fermion model.

Alternatively, one may investigate the finite temperature effect on the stability of Haldane-like phase. Physically, the topological state will still be protected by the bulk gap unless the thermal fluctuation overwhelms such gap. At the same time, the elevated temperature
has the potential to destroy Kondo screening and it is expected that we can detect the breakdown of the surface Kondo effect with the help of finite-temperature determinant quantum Monte Carlo algorithm widely used in the simulation of correlated electron problems.\cite{Alexandrov2015,Blankenbecler1981,Hirsch1985,Santo2003}
Future work in this direction will be of great help in clarifying the possibility of the temperature-driven surface Kondo breakdown mechanism.

\section{Acknowledgments}
This research was supported in part by NSFC under Grant No.~$11325417$, No.~$11674139$ and No.~$11504061$, the Fundamental Research Funds for the
Central Universities, Science Challenge Project No.~JCKY$2016212A502$ and the Foundation of LCP.

\section{Contribution statement}
Y. Zhong suggested the issue and carried out the calculation. All of authors wrote and revised this article.

\end{document}